\documentclass{elsart4}

\usepackage{graphicx}
\usepackage{amssymb}

\begin{document}
\begin{frontmatter}



\title{Two geometrically frustrated magnets studied by neutron diffraction.}

%
%
%
%
%
%

\author[llb]{I. Mirebeau\corauthref{1},}
\author[llb]{A. Apetrei,}
\author[llb]{I.N. Goncharenko,}
\author[ens]{R. Moessner}
%

\address[llb]{Laboratoire L\'eon Brillouin, CEA/CNRS UMR12, Centre d'Etudes de Saclay, 91191 Gif sur Yvette, France}
\address[ens]{Laboratoire de Physique th\'eorique de l'Ecole Normale Sup\'erieure, CNRS-UMR8549, 75005 Paris, France}
%
%
%
%


%
%
%
%

\corauth[1]{Corresponding Author Email :
mirebea@dsm-mail.saclay.cea.fr}


\begin{abstract}
In the pyrochlore compounds, Tb$_2$Ti$_2$O$_7$ and
Tb$_2$Sn$_2$O$_7$, only the Tb$^{3+}$ ions are magnetic. They
exhibit quite abnormal -- and, in view of their chemical
similarity, strikingly different -- magnetic behaviour, as probed
by neutron diffraction at ambient and applied pressure.
Tb$_2$Ti$_2$O$_7$ is a cooperative paramagnet (`spin liquid'),
without long range order at ambient pressure; however, it does
become ordered under pressure. By contrast, Tb$_2$Sn$_2$O$_7$
enters an ``ordered spin ice" state already at ambient pressure. We
analyse a simple model which already clearly exhibits some of the
qualitative features observed experimentally. Overall, comparing
these two compounds emphasizes the power of small perturbations in
selecting low-temperature states in geometrically frustrated
systems.
\end{abstract}

%
%



\end{frontmatter}

%
%
%
%
%
\section{Introduction}
Geometrical frustration (GF) \cite{hfmrev} is now widely studied
in solid state physics, as it seems to play a key role in original
phenomena recently observed in new materials. Examples include the
large anomalous Hall effect in ferromagnetic pyrochlores or
spinels \cite{Taguchi01}, the unconventional superconductivity
observed in water substituted Na$_x$CoO$_2$ with triangular Co
sheets \cite{Takada03}, or the interaction between electric and
magnetic properties of multiferroics materials \cite{Blake05}.

What is geometrical frustration? Most simply, it occurs when the
specific geometry of the lattice prevents magnetic interactions
from being satisfied simultaneously. In insulating systems such as
the rare earth pyrochlores, the impossibility of a simple N\'eel
ground state due to GF offers the possibility of finding a large
variety of alternative, magnetic and non-magnetic, short- or
long-ranged ordered states. In the most extreme case, paramagnetic
behaviour persists down to the lowest temperatures, leading to an
extended cooperative paramagnetic, or spin liquid, regime, in
which
 only short-range correlations result \cite{Villain79}.

Ferromagnetic interactions on the pyrochlore lattice may also be
frustrated, namely when the exchange is dominated by a strong
anisotropy which forces the spins in a tetrahedron to point along
their local, non-collinear easy axes \cite{Harris97}. This leads to the spin ice
state, whose degeneracy can be mapped onto that of real ice
\cite{Harris97}, leading to approximately the same entropy in the
ground state \cite{Ramirez99}.

In real compounds, the eventual choice of the stable magnetic
state depends on a subtle energy balance between the frustrated
first neighbour exchange energy term and perturbation terms of
various origins (longer range interactions, anisotropies,
quantum fluctuations, ...).  It is of course also determined by
thermodynamic parameters, such as temperature, pressure or
magnetic field. Counterintuitively, thermal fluctuations can even
induce order when ordered states permit softer fluctuations than
generic disordered ones. This effect is known as order by disorder
\cite{villain-shender} and is commonly encountered in frustrated magnetism. It has been well
studied by Monte-Carlo simulations, and also received some experimental confirmation \cite{Gukasov88-Champion03}. Pressure can change the
nature of, and balance between different terms in the Hamiltonian,
as they can depend on interatomic distances in different ways. An
applied field adds Zeeman energy, and can, for example, stabilize a subset of
the original ground states, at times resulting in magnetization
plateaus.


In this paper, we study a well known pyrochlore Tb$_2$Ti$_2$O$_7$,
which we investigated by neutron diffraction under extreme
conditions of temperature (down to 0.1K) and applied pressure (up
to 8.7 GPa). We review one of its most fascinating properties,
namely its ability to ``crystallize" or order magnetically under
pressure and we propose a new theoretical approach which accounts
for some important peculiarities of this effect. We also compare
Tb$_2$Ti$_2$O$_7$ to its sibling compound Tb$_2$Sn$_2$O$_7$, very
recently studied, which behaves as an ``ordered spin ice". Both
compounds have a fully chemically ordered structure, the
pyrochlore structure of cubic Fd$\overline{3}$m space group, where
the Tb$^{3+}$ magnetic ions occupy a GF network of corner sharing
tetrahedra. Although they differ only by the nature of the non-magnetic ion (Ti/Sn),
they show very different magnetic ground
states. The comparison sheds some light on how to select the
ground state through very small perturbations, one of the most
prominent characteristics of geometrical frustration.

\section{Tb$_2$Ti$_2$O$_7$ : a spin liquid orders under applied pressure}
Tb$_2$Ti$_2$O$_7$ is a famous example of a spin liquid,
investigated by numerous groups, where short range correlated Tb
spins fluctuate down to 70 mK at least \cite{Gardner99}, that is
more than 300 times below the typical energy scale of the magnetic
interactions (the Curie-Weiss constant $\theta_{\rm CW}$ of -19 K,
where the minus sign corresponds to AF interactions). The
persistence of these fluctuations was checked by muon relaxation
\cite{Gardner99}, at the time scale of the muon probe of about
$10^{-6}$ s. At shorter time scales, inelastic neutron scattering
showed a quasi elastic signal, whose energy linewidth strongly
decreases below about 1 K, indicating a stronger slowing down in
this temperature range \cite{Yasui02}. Coexisting with the spin liquid
phase, spin glass like irreversibilities and anomalies of the
specific heat were recently observed in the range 0.1 K-0.8 K
\cite{Hamaguchi04}. Using high pressure powder neutron diffraction
 \cite{Goncharenko04}, we observed two interesting phenomena
induced by pressure \cite{Mirebeau02} i) the onset of
antiferromagnetic long range order below a N\'eel temperature
T$_{\rm N}$ of about 2 K: ii) the enhancement of the magnetic
correlations in the spin liquid phase above T$_{\rm N}$. Just
below T$_{\rm N}$, the ordered phase coexists with the spin liquid
in a mixed solid-liquid phase, whose relative contributions vary
with pressure and temperature. The magnetic Bragg peaks of the
simple cubic lattice can be indexed from the crystal structure
 of Fd$\overline{3}$m symmetry, taking a propagation vector {\bf k}=(1,0,0). It
means that in the cubic unit cell with four Tb tetrahedra, two
tetrahedra are identical and two have reversed moment directions.
A longer wavelength modulation of this structure, involving a much
larger unit cell, was also observed in the powder data.

\begin{figure}
\centerline{\includegraphics[width=\columnwidth]{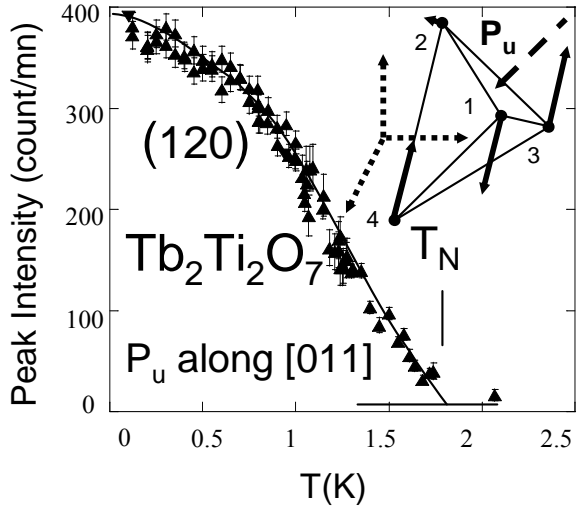}}
\caption{Tb$_2$Ti$_2$O$_7$ : an
 antiferromagnetic ordered state with {\bf k}=(1,0,0) propagation vector is
induced under pressure. Here an isotropic pressure P$_i$=2.0 GPa
is combined with uniaxial pressure P$_u$ =0.3 GPa along $[$011$]$ axis.
The variation of the peak intensity of the magnetic Bragg peak
(120) shows the N\'eel temperature. The local spin structure in a
tetrahedron has non compensated magnetization.} \label{fig:Fig1}
\end{figure}

 What is the pressure induced ground state? More fundamentally what is the role of pressure?
 To answer these questions, we performed single crystal neutron diffraction down to very
 low temperatures (0.14 K), combining hydrostatic pressure with anisotropic stress \cite{Mirebeau04}.
 We showed that both components play a role in inducing the long range order,
 and that the ordered moment and N\'eel temperature can be tuned
 by the direction of the stress.
  A stress along  a $[$110$]$ axis, namely along  the direction of the first neighbour distances between Tb$^{3+}$
  ions,
 is the most efficient in inducing magnetic order (Fig.~\ref{fig:Fig1}).

 \begin{table}[b]
     \centering
     \begin{tabular}{ccccc}
  \hline
  site & $x$ & $y$ & $z$ & $M$\\
  \hline
  1 & 0.5 & 0.5 & 0.5 & $[$1 0 -1$]$ \\
  2 & 0.25 & 0.25 & 0.5 & $[$1 0 1$]$ \\
  3 & 0.25 & 0.5 & 0.25& $[$-1 0 1$]$ \\
  4 & 0.5 & 0.25 & 0.25& $[$-1 0 1$]$ \\
  \hline
     \end{tabular}
     \caption{
Orientation of the magnetic moments in one tetrahedron in the
pressure induced state of Fig 1., deduced from the re
inement
of the magnetic structure. The stress component is along $[$0 1
1$]$. The atomic coordinates $x$, $y$, $z$, are expressed in the cubic
unit cell containing 4 tetrahedra. Two tetrahedra are identical
and two have reversed spin orientations.}\label{Table}
 \end{table}

FullProf refinements of the single crystal data allowed a determination of
the magnetic structure with better precision, especially the local spin structure within
a Tb tetrahedron. The structure corresponding to the best refinement (R$_F$=14\% is given
in table~\ref{Table}.  The bond 1-4 along the axis of the stress, which should be reinforced, has AF collinear spins. This corresponds to a natural expectation for AF first neighbor exchange.
The orientation of the spin 2 (orthogonal to the 3 others) is more surprising since with 3 collinear spins,
the exchange field on the fourth one should be also
collinear. Since no collinear structure gave a good fit
to the data, it suggests that the real spin structure
may be even more complex than the proposed one. In
any case, both powder and single crystal data yield an
important conclusion: we found that inside a tetrahedron, the magnetization is not compensated, namely
the vectorial sum of the four spins is non zero, (although it is of course compensated within the cubic
cell,  since magnetisations of the four tetrahedra
cancel two by two). This means that in the pressure induced ground
state, the local order does not correspond to any configuration
which mimimizes the energy in the spin liquid phase. In other
words, pressure does not select any energy state among those
belonging to the ground state degeneracy of the spin liquid (the
ground state expected if one considers Heisenberg spins coupled
via first neighbour AF exchange interactions only). The
anisotropic component (stress) relieves the frustration in a more
drastic way, by creating uncompensated bonds, associated with a
very small distortion of the pyrochlore lattice. In addition the
isotropic component shortens all distances in the same way,
increasing the frustrated exchange interaction. This effect could
also contribute to the increase of T$_{\rm N}$.

\begin{figure}
\centerline{\includegraphics[angle=0,width=\columnwidth]{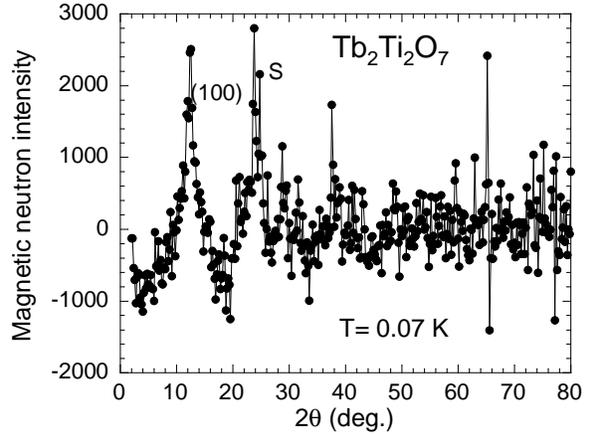}}
\caption{Tb$_2$Ti$_2$O$_7$ : a
mesoscopic order is induced by spontaneous strains at very low
temperature.
 Magnetic neutron diffraction spectra at 0.07 K, showing broad
 peaks close to the positions of the (100) magnetic peak and
 secondary magnetic phase (S) of the pressure-induced state
 \cite{Mirebeau02}.
 The spectrum of the spin liquid regime at 1.2 K has been subtracted.
 The incident neutron wavelength is 2.52 $\AA$.
 The broad magnetic peaks disappear at about 0.3 K.} \label{fig:Fig2}
\end{figure}

The prominent role of stress in inducing magnetic order raises a
subsequent question. Could it be stabilized  spontaneously by
internal stresses? To answer this question, we have now checked
the magnetic order at ambient pressure by neutron diffraction in
two Tb$_2$Ti$_2$O$_7$ samples with different heat treatments, down
to about 0.1 K. In an ``as cast" powder sample, we observe at 0.07
K  broad magnetic peaks close to the positions expected for the
pressure induced magnetic order (Fig.~\ref{fig:Fig2}). The
Lorentzian lineshape corresponds to a finite correlation length of
about 25 $\AA$ (2-3 cubic cells). The peaks disappear around 0.3
K. We also studied a single crystal, which was annealed at 1150
$^{o}$C for 25 hours to relieve internal stresses. In the second
case, the mesoscopic magnetic order is absent and only the liquid-like
correlations are observed, down to the minimum temperature of
0.15 K. Since both samples are chemically ordered and
stoichiometric within the accuracy of neutron diffraction, it
means that the mesoscopic order is induced by internal stresses.
The onset of this mesoscopic order may strongly influence the spin
glass irreversibilities and anomalies of the specific heat
observed in the same temperature range\cite{Hamaguchi04}, which
seem to depend on the heat treatment.

\section{Tb$_2$Sn$_2$O$_7$: an ordered spin ice state}
\begin{figure}
\centerline{\includegraphics[angle=0, width=2.7in]{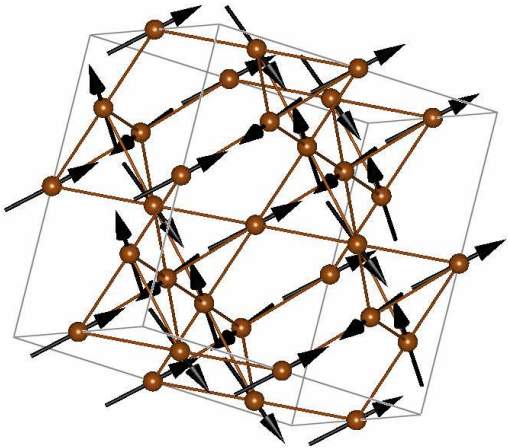}}
\caption{Tb$_2$Sn$_2$O$_7$ : an
 ordered spin ice structure: the local spin structure in a tetrahedron is close
 to the ``2 in-2 out" structure of  a spin ice, but individual tetrahedra are identical,
 leading to an ordered structure with {\bf k}=0 propagation vector and ferromagnetic character.} \label{fig:Fig3}
\end{figure}

In contrast to Tb$_2$Ti$_2$O$_7$,  Tb$_2$Sn$_2$O$_7$ undergoes a
transition to an ordered state already at ambient pressure.
The magnetic structure, very recently determined by powder neutron
diffraction experiments \cite{Mirebeau05} was called an ``ordered
spin ice".
 The local order within one tetrahedron is close to the
``two in-two out"
 configuration of spin ice, taking into account
 a small deviation of $13^{o}$. of the magnetic moments
 with respect to the local $<$111$>$ easy anisotropy axes.
In the canonical spin ice state,
 individual tetrahedra keep the mutual orientational disorder
 allowed by the ``ice rules", leading to short range order and ground state entropy \cite{Bramwell01}.
 Here the four tetrahedra of the unit cell are identical,
 leading to an ordered structure with {\bf k}=0 propagation vector (Fig.~\ref{fig:Fig3}). The resulting magnetic structure is non-collinear, but
 exhibits a ferromagnetic component, which represents about 37\%
 of the Tb$^{3+}$ ordered moment. This explains the ferromagnetic
 character of the transition, previously observed by magnetization \cite{Matsuhira02}.

 Together with the non-collinear magnetic structure, the original
effects of the frustration persist in the ordered phase of
Tb$_2$Sn$_2$O$_7$.  The magnetic order is stabilized in two steps
(1.3 K and 0.87 K) corresponding to anomalies of the specific
heat, and not in a classical second order transition. The
correlation length increases throughout the transition region, and
remains limited to 180 $\AA$  even at very low temperature.  The
ordered state coexists with slow collective fluctuations, in the
time scale of $10^{-4}$-$10^{-5}$ s. They were probed by comparing
the Tb$^{3+}$ moment value of 5.9(1) $\mu_B$ deduced from neutron
diffraction to the much lower value of 3.3(3) $\mu_B$ deduced from
the specific heat.

 The magnetic order in Tb$_2$Sn$_2$O$_7$ may be
 compared to that found by Champion {\it et al.} \cite{Champion02}, who considered
 the competition between
 first neighbor exchange and uniaxial anisotropy in a pyrochlore {\it ferromagnet}.
  The model involves two
parameters, the strength of the ferromagnetic interaction J and
that of the uniaxial anisotropy D$_{\rm a}$ along $<$111$>$ axes.
Ferromagnetic and spin ice states correspond to the cases D$_{\rm
a}$ /J = 0 and D$_{\rm a}$/ J = $\infty$, respectively. For finite
D$_{\rm a}$/J values, the magnetic order shows  many similarities
with the observed one. Namely: i) the ground state is ordered  in a
{\bf k}=0 four sublattice structure. ii) the local order within one
tetrahedron may also be deduced from the
spin ice structure. 
iii) the magnetic transition is of first order, changing to second
order with decreasing D$_{\rm a}$/J.

However, the deviations from the local spin ice structure are
different in the model and in the real system. In the model, spins
are uniformly canted towards the ferromagnetic 
direction. The ground state magnetization relative to the local
moment increases from 0.578=1/$\sqrt 3$ (the average
magnetization of a tetrahedron in the spin ice case) to 1 (the
ferromagnetic case) with decreasing D$_{\rm a}$/J. By contrast, in
 Tb$_2$Sn$_2$O$_7$, the deviations of the magnetic moments from the local $<$111$>$ axes
 actually reduce
 the magnetization (to about 0.37 in relative units). So the
 deviations of the magnetic moments from the local spin
 ice structure act in an opposite way to that predicted by the
 finite anisotropy ferromagnetic model.

 Finally, in Tb$_2$Sn$_2$O$_7$, the neutron and magnetic data together with the
 comparison with theory, suggest that here the
 effective first neighbor interaction becomes
 ferromagnetic, although the physics of the system cannot be
 simply reduced to the energy scheme assumed in ref. \cite{Champion02}.

\section{A simple model for the influence of stress}
An analysis of the effect of pressure applied to an individual
tetrahedron already manifests qualitatively two basic experimental
results observed in Tb$_2$Ti$_2$O$_7$: the much stronger influence
of an anisotropic stress along  a $[$110$]$ axis (as opposed to
hydrostatic pressure or to a stress along the
$[$100$]$ direction), and the presence of an uncompensated
magnetisation.

\begin{figure}
\centerline{\includegraphics[angle=0,
width=2.in]{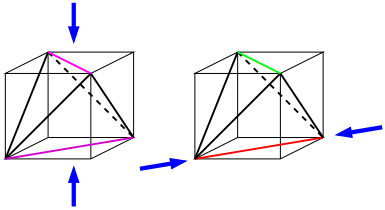}} \caption{Tetrahedron under uniaxial stress, denoted by the arrows. Left (right) panel: stress applied in the $[$100$]$ ($[$110$]$) direction. This splits the six
bonds into the following symmetry-inequivalent groups: the
bond along the $[$110$]$ direction (bottom), the one
perpendicular to it (top) -- which remain equivalent for $[$100$]$ but not for $[$110$]$ stress -- and the four remaining ones.} \label{fig:unitet}
\end{figure}

In the isotropic problem, all six bonds of the tetrahedron are
equivalent. Application of stress in the $[$110$]$ direction lowers this symmetry,
as shown in Fig.~\ref{fig:unitet}.
In symmetry terms, the bonds form a six-dimensional representation
of the tetrahedral group $T_d$, which decomposes into three
irreducible representations.  A singlet, A amounts to a
uniform change of all bonds together. Furthermore, there is a doublet,
E, the components of which correspond (a) to strengthening two
opposite bonds, and weakening the four others (or \textit{vice versa}) or (b) to weakening
an opposite pair of those four bonds, and strengthening the other
pair. Finally, each component of the triplet, T, implies a
strengthening/weakening (by an equal amount) of an opposite pair
of bonds \cite{pyroelastic}.

The crucial point is that the uniaxial  $[$110$]$ stress can
couple to all three representations. In its presence, there are three
(instead of only one) symmetry-inequivalent bond strengths
(see Fig.~\ref{fig:unitet}). In other
words, the Hamiltonian including the uniaxial $[$110$]$ stress has
a lower symmetry than the isotropic one.  The case of $[$1 0 0$]$ pressure is intermediate:
here only two representations are present, the triplet being absent, as illustrated in the left panel of
Fig.~\ref{fig:unitet}.

Whereas the initial
degeneracy of the isotropic system is a signature of the different
possible compromises between which bonds to frustrate and which to
satisfy, some of these choices have become forbidden as it is not
possible to trade off inequivalent bonds against one another.

For simplicity, let us consider a classical, isotropic Heisenberg
antiferromagnet at T=0 under stress. We have considered the spin
configurations which minimize the energy for different
orientations of the stress. We find that these configurations have
a compensated magnetization for a stress along $[$100$]$. By contrast, a
non-compensated magnetic moment can arise for a stress along a $[$110$]$ axis.
A summary of
the calculation is given below.

The results for an ice-type model
(i.e.\ a ferromagnet in the presence of anisotropy D$_{\rm a}$)
can be obtained along similar lines.  It needs to be borne in mind, however, that (a)
the strict ice model (D$_{\rm a}$/J=$\infty$) does not permit small
deviations of the spins from their preferred axes, and that (b) a ferromagnet will generically exhibit a non-compensated moment even in the absence of stress. By contrast, an anisotropic (but strain-free) antiferromagnet has a momentless ground state, which, however, is the simple FeF$_3$-type `all-in' or `all-out' ground state.

This classical isotropic Hamiltonian has a continuous two-parameter family of
degenerate ground states in the isotropic case \cite{hfmrev}. This
degeneracy is reduced by the strain. For example, in the presence of
an E-distortion weakening the average strength of the top/bottom
pair of bonds with respect to the other two pairs, a collinear
state will be selected. In this state, each spin is parallel to
its partner at the other end of the coloured bond, and
antiparallel to the other pair of spins. The total spin of the
tetrahedron thus remains compensated at zero. For an $E$ distortion of the opposite sign, the top
(and bottom) pair of spins will be antialigned; for the full pyrochlore lattice, this generates decoupled chain states.
In contrast to this situation, the presence of a $T$ distortion does not change the
energy of the isotropic ground
states relative to one another to first order. This happens
because it couples to a difference in the expectation value of
opposing bonds, a difference which vanishes in the unperturbed
ground states.

However, in higher order, a difference in relative
bond strength can induce a difference between the
expectation values of the scalar product $S_i\cdot S_j$ across the
bottom and top bonds. Such a difference is equivalent to an uncompensated total moment
of the tetrahedron (as a tetrahedron with zero moment necessarily
has equal expectation values of $S_i \cdot S_j$ on opposite
bonds).  The ground states in the presence of stress are thus
close to -- but not a subset -- of those of the isotropic system.

\section{Discussion}
In this section we briefly comment about the relevance of the
above model to Tb$_2$Ti$_2$O$_7$, and then focus on the origin of
the differences between the two compounds.

 At ambient pressure, the fact that Tb$_2$Ti$_2$O$_7$  does not order but remains
`liquid' down to 70 mK, is still a challenge to theory.
Sophisticated calculations taking into account the crystal field
energy \cite{Gingras00} together with dipolar interactions,
predict an Ising like behavior for the  Tb ion moment in the the
ground state (with the moment reduced with respect to the free ion
value) and an effective AF first neighbor
interaction\cite{Kao04-Enjalran04}. These calculations predict at
ambient pressure a transition to an AF order similar to that
found in FeF$_3$ (a {\bf k}=0 structure with an ``all in-all out"
local configuration) below 1-2K, which is however not what is
observed experimentally.

The simple model discussed above, already proposed on a more empirical
basis in Ref. \cite{Mirebeau05}, reproduces the main characteristics of the
pressure-induced
 state in Tb$_2$Ti$_2$O$_7$, namely the stronger effect of degeneracy lifting of
 the $[$110$]$ over the $[$100$]$ stress, and the appearance of an uncompensated magnetisation in the former case.
 This is presumably the case because it incorporates the most fundamental
property of the stress, namely the explicit symmetry breaking it
induces. This effect should occur in qualitatively the same way in a
much larger class of models.

These results are therefore rather robust but, by the same token, they are also only
qualitative: the model in its current form yields little
information on the detailed Hamiltonian of the system, nor the
origin of the effective nearest-neighbour interaction J and of its
sensitivity to pressure. In particular, we have not been able to reproduce the detailed finite-temperature spin structure.


We now turn to Tb$_2$Sn$_2$O$_7$, which behaves as an ordered spin
ice. Neutron data as compared with theory strongly suggest that
the effective first neighbour interaction has now become
ferromagnetic. What is the reason for this change? We can propose
the following explanation. In the ``true" spin ices
(Ho$_2$Ti$_2$O$_7$ or Dy$_2$Ti$_2$O$_7$ with stronger uniaxial
anisotropy), it was shown that the effective ferromagnetic
interaction results from the influence of the dipolar coupling
which overcomes the weak
 AF superexchange \cite{Bramwell01}. Taking the same conventional
 notations, the effective first neighbour interaction J$_{eff}$ is expressed
 as J$_{eff}$= J$_{nn}$+D$_{nn}$, where J$_{nn}$=J/3 and D$_{nn}$=5D/3 are the
superexchange and dipolar energy scales, respectively. In
Tb$_2$Ti$_2$O$_7$ (J$_{nn}$=-0.88 K, D$_{nn}$=0.8 K from ref.
\cite{Kao04-Enjalran04}), this effective interaction remains AF.
In Tb$_2$Sn$_2$O$_7$, Sn substitution enlarges the unit cell (from
a= 10.149 to 10.426 $\AA$ in Ti and Sn compounds respectively).
This expansion $\Delta a/a \sim 2.7\%$, equivalent to a negative
chemical pressure of about 12-15 GPa, should strongly decrease the
AF superexchange interaction J. Assuming roughly a decrease of
J$_{nn}$ in the ratio of the Curie-Weiss constants (-19 K and -12
K in Ti and Sn compounds respectively) without big changes in the
dipolar constant, we get J$_{eff}$ = 0.18 K $>$ 0 for
Tb$_2$Sn$_2$O$_7$. Therefore the expansion in the unit cell
induced by Sn substitution might be enough to switch the compound
to the spin ice region of the phase diagram \cite{DenHertog00}.

  To go further, microscopic models should take
 into account the exact nature of the anisotropy, which is not simply uniaxial in the Tb compounds\cite{Gingras00}.
 This involves a reinvestigation of the crystal field
 levels, currently in progress. It could 
  exhibit more
 subtle differences between the two compounds than the simple effect of a chemical
 pressure discussed above.

 In conclusion, the two compounds studied here clearly show the rich variety of
  behaviour exhibited by geometrically frustrated magnets.
  Comparing them allows one to understand better the key role
  played by small perturbations in selecting one peculiar state among the many potential
  magnetic states.

We thank  A. Gukasov  and  O. Isnard for their help in the neutron
measurements at LLB and ILL respectively. We also thank G.
Dhalenne, A. Revcolevschi, A. Forget and D. Colson, who provided
the single crystal and powdered samples. R. M.  thanks S. Sondhi
and O. Tchernyshyov for collaboration on related work. He was
supported in part by the Minist\`ere de la Recherche with an ACI
grant.
%
%
%
%

%
%
%
%



\begin{thebibliography}{00}

\bibitem{hfmrev} For an introduction to frustrated magnets, see
R. Moessner, Can. J. Phys. \textbf{79}, 1283, (2001); 
reviews of exact diagonalizations and experiments, respectively,
are C. Lhuillier, P. Sindzingre and J.-B. Fouet, Can. J. Phys.
\textbf{79}, 1525, (2001) 
and P. Schiffer and A. P. Ramirez, Comments Cond. Mat. Phys.
\textbf{18}, 21, (1996). 



\bibitem{Taguchi01} Y. Taguchi, Y. Oohara, H. Yoshisawa, N. Nagaosa, Y. Tokura Science \textbf{291}, 2573
(2001).
\bibitem{Takada03} K. Takada {\em et al.} Nature \textbf{422}, 53, (2003).
\bibitem{Blake05} G. R. Blake {\em et al.}, Phys. Rev. B. \textbf{71}, 214402, (2005).
\bibitem{Villain79} J. Villain Z. Phys. B \textbf{33}, 31; (1979).
\bibitem{Harris97} M. J. Harris {\em et al.}, Phys. Rev. Lett. \textbf{79}, 2554 (1997).
\bibitem{Ramirez99} A. P. Ramirez, A Hayashi, R. J. Cava, R. Siddhartan, B.S. Shastry Nature, \textbf{399}, 333, (1999). 
\bibitem {villain-shender}
 J. Villain, R. Bidaux, J. P. Carton, R. Coute J. Phys. (Paris)
 \textbf{41}, 1263, (1980); E. F. Shender Sov. Phys. JETP \textbf{56}, 178,
 (1982).
 \bibitem{Gukasov88-Champion03} A. G. Gukasov {\em et al.}, Europhys. Lett. \textbf{79}, 2554 (1997); J. D. M. Champion {\em et al.}, Phys. Rev. B \textbf{68}, 020401 R
 (2003).
\bibitem{Gardner99} J. S. Gardner {\em et al.} Phys. Rev. Lett. \textbf{82}, 1012, (1999).
\bibitem{Yasui02} Y. Yasui {\em et al.}, J. Phys. Soc. Jpn. \textbf{71}, 599, (2002). 
\bibitem{Hamaguchi04} N. Hamaguchi, T. Matsushita, N. Wada, Y.
Yasui and S. Masatoshi, Phys. Rev. B \textbf{69}, 132413, (2004). 
\bibitem{Goncharenko04} I. N. Goncharenko, High Pressure Res. \textbf{24},193,(2004).
\bibitem{Mirebeau02} I. Mirebeau, I.N. Goncharenko, P.Cadavez-Peres, S. T. Bramwell, M.J.P. Gingras and J. S. Gardner
Nature  \textbf{420}, 54, (2002).
\bibitem{Mirebeau04} I. Mirebeau, I. N. Goncharenko, G. Dhalenne,
A. Revcolevschi, Phys. Rev. Lett. \textbf{93}, 187204, (2004); I.
Mirebeau and I. Goncharenko J. Phys. Cond. Mat. \textbf{17}, S771,
(2005).
\bibitem{Mirebeau05} I. Mirebeau {\em et al.} Phys. Rev. Lett.  \textbf{94}, 246402, (2005).
\bibitem{Bramwell01} S. T. Bramwell and M. J. P. Gingras Science {\bf 294}, 14, (2001).
\bibitem{Matsuhira02} K. Matsuhira {\em et al.} J. Phys. Soc. Jpn. \textbf{71},1576,(2002).
\bibitem{Champion02}
J. D. M. Champion, S. T. Bramwell, P. C. W. Holdsworth and M. J.
Harris, Europhys. Lett. {\bf 57}, 93 (2002).
\bibitem{pyroelastic}
Y. Yamashita and K. Ueda, Phys. Rev. Lett. {\bf 85}, 4960 (2000);
O. Tchernyshyov, R. Moessner, S. L. Sondhi, Phys. Rev. Lett. {\bf
88}, 067203 (2002).
\bibitem{Gingras00} M. J. P. Gingras {\em et al.} Phys. Rev. B {\bf 62}, 6496, (2000).
\bibitem{Kao04-Enjalran04} Y. Kao, M. Enjalran, A. Del Maestro, H. Molavian
and M. J. P. Gingras Phys. Rev. B {\bf 68}, 172407, (2002); M.
Enjalran and M. J. P. Gingras Phys. Rev. B {\bf 70}, 174426,
(2004).
\bibitem{DenHertog00} B. C. Den Hertog and M. J. P. Gingras Phys. Rev.
Lett. {\bf 84}, 3430, (2000).


\end{thebibliography}
\end{document}